\documentclass{article}

\usepackage{PRIMEarxiv}

\usepackage[utf8]{inputenc} 
\usepackage[T1]{fontenc}    
\usepackage{hyperref}       
\usepackage{url}            
\usepackage{booktabs}       
\usepackage{amsfonts}       
\usepackage{nicefrac}       
\usepackage{microtype}      
\usepackage{lipsum}
\usepackage{fancyhdr}       
\usepackage{graphicx}       
\usepackage{verbatim}
\usepackage{listings}
\usepackage{xcolor}

\usepackage{booktabs}
\usepackage{tabularx}
\usepackage{float} 
\usepackage{placeins} 
\graphicspath{{media/}}     

\definecolor{eclipseStrings}{RGB}{34,139,34}
\definecolor{eclipseKeywords}{RGB}{220,20,60}
\colorlet{numb}{magenta!60!black}
\definecolor{lightgray}{rgb}{0.95,0.95,0.95}

\lstdefinelanguage{json}{
    basicstyle=\tiny\ttfamily,
    commentstyle=\color{eclipseStrings}, 
    stringstyle=\color{eclipseKeywords}, 
    numbers=none,
    showstringspaces=false,
    breaklines=true,
    frame=lines,
    backgroundcolor=\color{lightgray}, 
    string=[s]{"}{"},
    comment=[l]{:\ "},
    morecomment=[l]{:"},
    literate=
        *{0}{{{\color{numb}0}}}{1}
         {1}{{{\color{numb}1}}}{1}
         {2}{{{\color{numb}2}}}{1}
         {3}{{{\color{numb}3}}}{1}
         {4}{{{\color{numb}4}}}{1}
         {5}{{{\color{numb}5}}}{1}
         {6}{{{\color{numb}6}}}{1}
         {7}{{{\color{numb}7}}}{1}
         {8}{{{\color{numb}8}}}{1}
         {9}{{{\color{numb}9}}}{1}
}

\title{Decoding MIE: A Novel Dataset Approach Using Topic Extraction and Affiliation Parsing
}

\author{
  Ehsan Bitaraf \\
 Rajaei Cardiovascular Reserach Institute \\
   Iran University of Medical Sciences \\
  Tehran, Iran\\
  \texttt{bitaraf.e@iums.ac.ir} \\
   \And
 Maryam Jafarpour \\
  Center for Medical Data Science \\
  Medical University of Vienna \\
  Vienna, Austria\\
  \texttt{maryam.jafarpour@meduniwien.ac.at} \\
}

\begin{document}
\maketitle

\begin{abstract}
The rapid expansion of medical informatics literature presents significant challenges in synthesizing and analyzing research trends. This study introduces a novel dataset derived from the Medical Informatics Europe (MIE) Conference proceedings, addressing the need for sophisticated analytical tools in the field. Utilizing the Triple-A software, we extracted and processed metadata and abstract from 4,606 articles published in the "Studies in Health Technology and Informatics" journal series, focusing on MIE conferences from 1996 onwards. Our methodology incorporated advanced techniques such as affiliation parsing using the TextRank algorithm. The resulting dataset, available in JSON format, offers a comprehensive view of bibliometric details, extracted topics, and standardized affiliation information. Analysis of this data revealed interesting patterns in Digital Object Identifier usage, citation trends, and authorship attribution across the years. Notably, we observed inconsistencies in author data and a brief period of linguistic diversity in publications. This dataset represents a significant contribution to the medical informatics community, enabling longitudinal studies of research trends, collaboration network analyses, and in-depth bibliometric investigations. By providing this enriched, structured resource spanning nearly three decades of conference proceedings, we aim to facilitate novel insights and advancements in the rapidly evolving field of medical informatics.
\end{abstract}


\section{Introduction}
In the rapidly evolving landscape of artificial intelligence (AI), the role of high-quality, diverse datasets cannot be overstated. These datasets serve as the foundation upon which AI models are built, trained, and refined, ultimately shaping the capabilities and performance of AI systems across various domains\cite{1,2}. The medical field, in particular, stands to benefit immensely from the application of AI, with potential improvements in diagnosis, treatment planning, and healthcare management \cite{3}.

The Medical Informatics Europe (MIE) Conference, a cornerstone event in the field of medical informatics, generates a wealth of valuable scientific content annually. However, the sheer volume of articles presented at such conferences poses a significant challenge for researchers and practitioners attempting to extract meaningful insights and identify emerging trends \cite{4}. This challenge underscores the need for sophisticated tools and methodologies to analyze and synthesize large volumes of scientific literature effectively \cite{5}.

In response to this need, we present a novel dataset derived from the MIE Conference proceedings, processed using advanced techniques such as topic extraction and affiliation parsing. This dataset not only exemplifies the power of open-source data in driving scientific progress but also demonstrates the potential of intelligent tools in managing and analyzing vast amounts of scientific information.

By making datasets freely available to the research community, we foster collaboration, encourage reproducibility, and accelerate the pace of innovation \cite{6}. Open-source datasets democratize access to valuable information, enabling researchers from diverse backgrounds to contribute to the advancement of AI and its applications in fields like medical informatics \cite{7}.

Furthermore, the application of smart tools for analyzing scientific articles across various disciplines has become increasingly crucial. As the volume of published research continues to grow exponentially, traditional manual methods of literature review and synthesis become increasingly untenable \cite{8}. Advanced techniques such as natural language processing, machine learning, and data mining offer powerful means to extract key insights, identify patterns, and uncover hidden relationships within large corpora of scientific literature \cite{8}.

This paper presents a comprehensive overview of our dataset preparation process. It details the methodologies employed in topic extraction and affiliation parsing. By providing this dataset and elaborating our approach, we aim to contribute to the broader scientific community's efforts in leveraging AI for knowledge discovery and synthesis in the field of medical informatics.

\section{Related Work}
\label{sec:headings}
In the realm of data-driven research, several notable datasets have significantly contributed to advancing various scientific fields. This section highlights key datasets that have set benchmarks in their respective domains, providing a context for the novel dataset approach used in our study.

\subsection{CORD-19}
The COVID-19 Open Research Dataset (CORD-19) is a comprehensive collection of over 200,000 research papers on COVID-19 curated by the Allen Institute for AI. This dataset has been instrumental in facilitating rapid scientific discovery and understanding of the COVID-19 pandemic by providing researchers with a rich repository of scholarly articles for text mining and data analysis. The extensive use of CORD-19 underscores the importance of well-curated datasets in addressing global health crises \cite{9}.
\subsection{S2ORC (Semantic Scholar Open Research Corpus)}
S2ORC is another significant dataset, consisting of a vast collection of research papers across various scientific disciplines. This open research corpus is designed to support large-scale text mining and natural language processing (NLP) tasks, enabling researchers to explore and analyze scientific literature comprehensively. S2ORC's broad scope and accessibility make it a valuable resource for advancing scientific research and innovation \cite{10}.
\subsection{ScisummNet}
ScisummNet is a specialized dataset focused on summarizing scientific papers. It provides annotated summaries of research articles, which are crucial for developing and evaluating automatic summarization algorithms. By offering a structured approach to summarizing scientific literature, ScisummNet aids in enhancing the efficiency and accuracy of information retrieval in academic research \cite{11}.
\subsection{TaeC}
The TaeC dataset is a manually annotated text dataset designed for trait and phenotype extraction and entity linking in wheat breeding literature. This dataset supports the agricultural research community by providing detailed annotations that facilitate the extraction of valuable information related to wheat traits and phenotypes, thereby advancing the field of plant breeding and genetics \cite{12}.
\subsection{MPI-LIT}
MPI-LIT is a literature-curated dataset of microbial binary protein-protein interactions. This dataset is beneficial for researchers studying microbial interactions, as it provides curated information on protein interactions, which is critical for understanding microbial functions and developing new biotechnological applications \cite{13}.
\section{Methodology}
In this study, we employed a structured and multi-stage approach to collect, process, and prepare a comprehensive dataset of articles from the Medical Informatics Europe (MIE) Conference. The primary tool used for this process was the Triple-A software \cite{14}, which facilitated data retrieval and preprocessing. This section outlines the various steps involved in constructing the dataset, including metadata acquisition, affiliation parsing, and topic extraction.
\subsection{Data Acquisition}
We began by sourcing the metadata for the articles from the PubMed database, utilizing the Triple-A tool to streamline the extraction process. Our initial query targeted all publications within the journal series "Studies in Health Technology and Informatics" (Stud Health Technol Inform), which regularly publishes MIE conference proceedings. This ensured we captured a wide range of articles, from which we could later refine the dataset.

To focus specifically on MIE conference papers, we filtered out irrelevant volumes that did not correspond to MIE-related events. This step ensured that the dataset was centered around contributions presented at the MIE Conference, maintaining a consistent scope for further analysis.

\begin{table}[H]
\centering
\caption{MIE Conferences (1996-2024)}
\begin{tabularx}{\textwidth}{>{\raggedright\arraybackslash}X >{\raggedright\arraybackslash}X >{\raggedright\arraybackslash}X p{5cm}}
\toprule
\textbf{Tag} & \textbf{Volume} & \textbf{Year} & \textbf{Place} \\
\midrule
MIE2024	&316	&2024 &Athens, Greece \\
MIE2023 & 302 & 2023 & Gothenburg, Sweden \\
MIE2022 & 294 & 2022 & Nice, France \\
MIE2021 & 281 & 2021 & Athens, Greece (e-Conference) \\
MIE2020 & 270 & 2020 & Geneva, Switzerland \\
MIE2018 & 247 & 2018 & Gothenburg, Sweden \\
MIE2017 & 235 & 2017 & Manchester, UK \\
MIE2016 & 228 & 2016 & Munich, Germany (@HEC2016) \\
MIE2015 & 210 & 2015 & Madrid, Spain \\
MIE2014 & 205 & 2014 & Istanbul, Turkey \\
MIE2012 & 180 & 2012 & Pisa, Italy \\
MIE2011 & 169 & 2011 & Oslo, Norway \\
MIE2009 & 150 & 2009 & Sarajevo, Bosnia-Herzegovina \\
MIE2008 & 136 & 2008 & Gothenburg, Sweden \\
MIE2006 & 124 & 2006 & Maastricht, Netherlands \\
MIE2005 & 116 & 2005 & Geneva, Switzerland \\
MIE2003 & 95 & 2003 & Sant Malo, France \\
MIE2002 & 90 & 2002 & Budapest, Hungary \\
MIE2000 & 77 & 2000 & Hanover, Germany \\
MIE1999 & 68 & 1999 & Ljubljana, Slovenia \\
MIE1997 & 43 & 1997 & Thessaloniki, Greece \\
MIE1996 & 34 & 1996 & Copenhagen, Denmark \\
\bottomrule
\end{tabularx}
\label{tab:mie-conferences}
\end{table}

\FloatBarrier  

\subsection{Affiliation Parsing}
To extract structured information from affiliations, we employed a specialized module available on GitHub\footnote[1]{\href{https://github.com/titipata/affiliation\_parser}{https://github.com/titipata/affiliation\_parser}}. This module utilizes a rule-based approach to parse affiliations into distinct components, including country, department, institution, location, zipcode, and email. 

The affiliation parser works by analyzing the text and identifying specific patterns and keywords associated with each component. It then categorizes the information accordingly, providing a structured representation of the affiliation data. This approach allows for more efficient organization and analysis of institutional information within the dataset.

\textbf{Key components extracted}:
\begin{itemize}
\item Country
\item Department
\item Institution
\item Location
\item Zipcode
\item Email
\end{itemize}
While the module offers robust parsing capabilities, it's important to note that the accuracy of the results can vary depending on the format and consistency of the input data. To mitigate potential errors, we implemented additional validation steps and manual checks for ambiguous cases.

The affiliation parser also provides an option to cross-reference the extracted institutions with the Global Research Identifier Database (GRID)\cite{15}. This feature enables verification and standardization of institution names, enhancing the overall quality and consistency of the affiliation data in our dataset.

By leveraging this affiliation parsing tool, we were able to transform unstructured affiliation strings into a structured format, facilitating more comprehensive analysis and insights into the institutional landscape represented in the MIE conference dataset.

\subsection{Topic Extraction}
For topic extraction, we applied the TextRank algorithm \cite{16}, a graph-based natural language processing (NLP) technique well-suited for unsupervised keyword and sentence extraction. TextRank operates by constructing a graph where the nodes represent words or sentences, and edges represent relationships between them based on co-occurrence. By identifying the most important nodes in this graph, TextRank effectively extracts key topics without the need for labeled training data. This algorithm was particularly beneficial for our dataset as it allowed us to extract relevant topics from the titles and abstracts of the articles spanning various languages and medical informatics domains, making the method adaptable and scalable across different contexts.
\subsection{Data Output and Formatting}
The final dataset was prepared in multiple formats to maximize usability and accessibility for a broad range of users. The primary output was structured in JSON format, allowing for flexibility in data handling and compatibility with various applications. 
\subsection{Dataset Maintenance and Update Protocol}

To ensure the longevity and relevance of the MIE conference dataset, we developed a systematic, reproducible update protocol implemented through a GitHub repository\footnote[2]{\href{https://github.com/EhsanBitaraf/dataset-mie-literature}{https://github.com/EhsanBitaraf/dataset-mie-literature}}. This protocol comprises five sequential steps, designed to be executed periodically to incorporate new conference proceedings:  
\begin{enumerate}
\item Configuration Validation (step01\_check\_config): This initial step verifies the integrity and accuracy of the TripleA program settings, ensuring consistency in data retrieval and processing parameters across update cycles.

\item PubMed Data Acquisition (step02\_get\_pubmed): Utilizing the PubMed API, this step retrieves the PubMed Identifiers (PMIDs) for all articles published in the "Studies in Health Technology and Informatics" journal series. This comprehensive approach ensures capture of all potential MIE conference papers.

\item Data Processing Pipeline (step03\_run\_pipline): This step executes a series of data preparation, topic extraction, and affiliation parsing operations. It applies our established methodologies for information enrichment, maintaining consistency with the original dataset creation process.

\item Conference-Specific Filtering (step04\_extract\_volume): To isolate MIE conference proceedings from other publications in the journal series, this step filters the collected data based on volume numbers associated with MIE conferences. This selective approach ensures the dataset's focus and relevance.

\item Dataset Compilation (step05\_create\_dataset): The final step aggregates the processed and filtered data into the standardized dataset structure, ready for integration with the existing dataset or for standalone analysis of new proceedings.
\end{enumerate}
This modular, step-wise approach facilitates transparency, allows for targeted modifications if required, and ensures the dataset remains current with ongoing MIE conferences. The GitHub repository serves as both a version control system and a collaborative platform for potential community contributions to dataset maintenance.
\section{Value of the Data}
The Medical Informatics Europe (MIE) Conference dataset presented in this article offers significant value to researchers, practitioners, and policymakers in the field of medical informatics. The following points highlight the key benefits and potential applications of this dataset:

\begin{enumerate}
   \item \textbf{Longitudinal Analysis of Medical Informatics Trends}:
   \begin{itemize}
     \item The dataset, covering MIE conferences from 1997 onwards, enables researchers to track the evolution of medical informatics over three decades.
     \item Temporal analysis can reveal shifting focus areas, emerging topics, and the lifecycle of various subfields within medical informatics.
   \end{itemize}
   \item \textbf{Topic Modeling and Content Analysis}:
\begin{itemize}
\item The inclusion of extracted topics for each article facilitates in-depth content analysis without the need for full-text processing.
\item Researchers can identify popular themes, track the emergence of new concepts, and analyze the interconnections between different areas of medical informatics.
\end{itemize}
\item \textbf{Bibliometric Studies}:
\begin{itemize}
\item Citation counts and publication details allow for comprehensive bibliometric analyses.
\item Researchers can identify influential papers, authors, and institutions in the field of medical informatics.
\item The data supports studies on citation patterns, impact assessment, and the diffusion of ideas within the community.
\end{itemize}
\item \textbf{Collaboration Network Analysis}:
\begin{itemize}
\item Author information and affiliation data enable the construction and analysis of collaboration networks.
\item Researchers can study patterns of international collaboration, institutional partnerships, and the role of key individuals in shaping the field.
\end{itemize}
\item \textbf{Geographical Distribution of Research}:
\begin{itemize}
\item The affiliation parsing results, particularly the country and institution information, allow for geographical analysis of medical informatics research.
\item This data can inform policy decisions related to research funding, educational programs, and international collaborations.
\end{itemize}
\item \textbf{Curriculum Development and Education}:
\begin{itemize}
\item The dataset provides a comprehensive overview of topics in medical informatics, which can be valuable for developing or updating educational curricula.
\item Educators can use the data to ensure that their teaching reflects current trends and important historical developments in the field.
\end{itemize}
\item \textbf{Industry and Technology Tracking}:
\begin{itemize}
\item By analyzing the topics and keywords over time, the dataset can provide insights into the adoption and impact of various technologies in healthcare informatics.
\item This information is valuable for industry stakeholders, technology developers, and healthcare organizations planning IT investments.
\end{itemize}
\item \textbf{Research Gap Identification}:
\begin{itemize}
\item By mapping the landscape of published work, researchers can identify underexplored areas or topics that warrant further investigation.
\item This can guide future research directions and funding allocations in the field of medical informatics.
\end{itemize}
\item \textbf{Conference Impact Assessment}:
\begin{itemize}
\item The dataset allows for analysis of the MIE conference's impact on the field of medical informatics over time.
\item Organizers and sponsors can use this information to assess the conference's role in shaping the discipline and to plan future events.
\end{itemize}
\item \textbf{Natural Language Processing (NLP) and Machine Learning Applications}:
\begin{itemize}
\item The structured nature of the dataset, with extracted topics and keywords, makes it an excellent resource for training and testing NLP and machine learning models focused on medical informatics literature.
\end{itemize}
\item \textbf{Interdisciplinary Research Opportunities}:
\begin{itemize}
\item The dataset can facilitate studies at the intersection of medical informatics with other fields such as computer science, healthcare management, and public health.
\item Researchers from various disciplines can use this data to understand how their field intersects with medical informatics.
\end{itemize}
\item \textbf{Policy and Funding Impact Analysis}:
\begin{itemize}
\item By correlating the dataset with information on research funding and policy initiatives, analysts can study the impact of various interventions on the field of medical informatics.
\end{itemize}
\end{enumerate}
This dataset represents a valuable resource for the medical informatics community and related fields. Its comprehensive nature, combining bibliometric data with advanced topic extraction and affiliation parsing, provides a unique opportunity for multifaceted analyses. As the field of medical informatics continues to evolve rapidly, this dataset offers a solid foundation for understanding its historical development and current state, while also providing insights that can shape its future direction.
\section{Dataset Description}
The dataset described in this article comprises a single JSON file containing detailed information about articles presented at the Medical Informatics Europe (MIE) Conference. This dataset was collected using the Triple-A tool and further enhanced through advanced techniques such as topic extraction and affiliation parsing.
\subsection{File Structure}
The JSON file contains a list of objects, where each object represents a single article and includes various fields of information. Below is an example of the structure for a single article entry:

\begin{lstlisting}[language=json]
[
    {
        "title": "Validation of Multiple Path Translation for SNOMED CT Localisation.",
        "year": "2022",
        "journal_issn": "1879-8365",
        "language": "eng",
        "abstract": "The MTP (multiple translation paths) approach supports human translators in clinical terminology localization. It exploits the results of web-based machine translation tools and generates, for a chosen target language, a scored output of translation candidates for each input terminology code. We present first results of a validation, using four SNOMED CT benchmarks and three translation engines. For German as target language, there was a significant advantage of MTP as a generator of plausible translation candidate lists, and a moderate advantage of the top-ranked MTP translation candidate over single best performing direct-translation approaches.",
        "doi": "10.3233/SHTI220641",
        "pmid": "35612259",
        "citation_count": 0,
        "IOSPressVolume": "294",
        "publication_type": [
            "Journal Article"
        ],
        "authors": [
            "Schulz, S",
            "Boeker, M",
            "Prunotto, A"
        ],
        "keywords": [
            "Ethnicity",
            "Humans",
            "Language",
            "Systematized Nomenclature of Medicine",
            "Translations",
            "Machine Translation",
            "NLP",
            "SNOMED CT"
        ],
        "topics": [
            "translation candidates",
            "plausible translation candidate lists",
            "clinical terminology localization",
            "multiple translation paths",
            "human translators",
            "SNOMED CT Localisation",
            "target language",
            "approach",
            "web-based machine translation tools",
            "single best performing direct-translation approaches"
        ],
        "affiliation_countries": [
            "austria",
            "germany"
        ],
        "affiliations": [
            "IMI, Medical University of Graz, Austria.",
            "Institute for AI in Healthcare, Technical University of Munich, Germany."
        ]
    }
]
\end{lstlisting}

\subsection{Field Descriptions}
Each article entry in the dataset contains the following fields:
\begin{table}[H] 
\centering
\caption{List of dataset fields}
\label{tab:dataset-fields}
\begin{tabularx}{\textwidth}{>{\raggedright\arraybackslash}X >{\raggedright\arraybackslash}p{7cm}}
\toprule
\textbf{Field} & \textbf{Description} \\
\midrule
title & The full title of the article \\
\addlinespace
year & The publication year of the article \\
\addlinespace
abstract & The abstract of the paper \\
\addlinespace
journal\_issn & The International Standard Serial Number (ISSN) of the journal \\
\addlinespace
language & The language of the article (e.g., "eng" for English) \\
\addlinespace
doi & The Digital Object Identifier of the article, if available \\
\addlinespace
pmid & The PubMed ID of the article \\
\addlinespace
citation\_count & The number of times the article has been cited \\
\addlinespace
IOSPressVolume & The volume number in the IOS Press publication series \\
\addlinespace
publication\_type & The type of publication (e.g., "Journal Article") \\
\addlinespace
authors & A list of the authors' names \\
\addlinespace
keywords & A list of keywords associated with the article in PubMed \\
\addlinespace
topics & The first 10 topics extracted through an unsupervised topic extraction mechanism \\
\addlinespace
affiliation\_countries & The countries associated with the authors' affiliations, extracted using the described algorithm \\
\addlinespace
affiliations & The list of authors' affiliations \\
\bottomrule
\end{tabularx}
\end{table}
\FloatBarrier  
\subsection{Data Processing and Enhancement}
This dataset has been enriched through several advanced techniques:
\begin{enumerate}
\item \textbf{Topic Extraction}: An unsupervised topic extraction mechanism was applied to identify the most relevant topics for each article. The 'topics' field lists the top 10 extracted topics, providing a quick overview of the article's content without relying solely on author-provided keywords.
\item \textbf{Affiliation Parsing}: An algorithm was developed to extract and standardize information about authors' affiliations. This process resulted in the 'affiliation\_coutries' field, which provides valuable insights into the geographical and institutional distribution of research in medical informatics.
\item \textbf{Bibliometric Data}: The inclusion of citation counts and publication details allows for bibliometric analyses, enabling researchers to identify influential papers and trends in the field.
\end{enumerate}
This comprehensive dataset offers a rich resource for researchers interested in the evolution of medical informatics, collaboration patterns, and trending topics in the field. The combination of bibliometric data, extracted topics, and standardized affiliation information provides multiple avenues for analysis and insights into the MIE conference contributions over time.
\section{Results}
The analysis of the Medical Informatics Europe (MIE) Conference dataset, prepared using the Triple-A tool and enhanced with topic extraction and affiliation parsing techniques, revealed several interesting patterns and characteristics.
\subsection{Dataset Composition}
The dataset comprises a total of 4,606 articles from the MIE Conference proceedings. This substantial corpus provides a comprehensive view of the research trends and developments in medical informatics over the years covered by the dataset.
\subsection{Digital Object Identifier (DOI) Assignment}
\begin{figure}[H]
  \centering
  \caption{Articles with and without DOI by Year}
\includegraphics[width=0.8\textwidth]{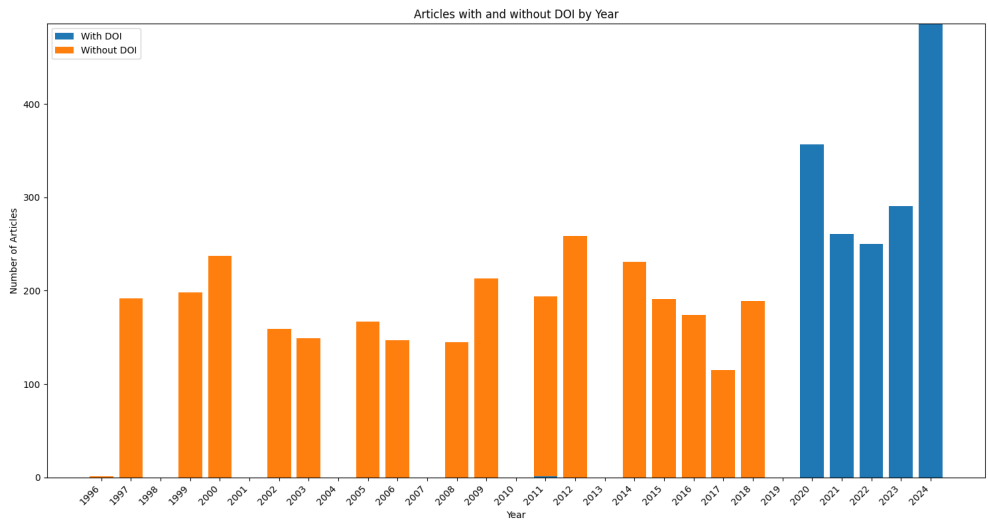}
  \label{fig:fig1}
\end{figure}

A notable trend in the dataset is the inconsistent assignment of Digital Object Identifiers (DOIs) across the years. As illustrated in Figure \ref{fig:fig1}, only articles from the most recent three years consistently included DOIs in their metadata. This observation highlights the evolving practices in digital article identification and the gradual adoption of DOIs in the field of medical informatics.
\subsection{Citation Patterns}
\begin{figure}[H]
  \centering
  \caption{Articles with No Citations vs At Least One Citation by Year}
\includegraphics[width=0.8\textwidth]{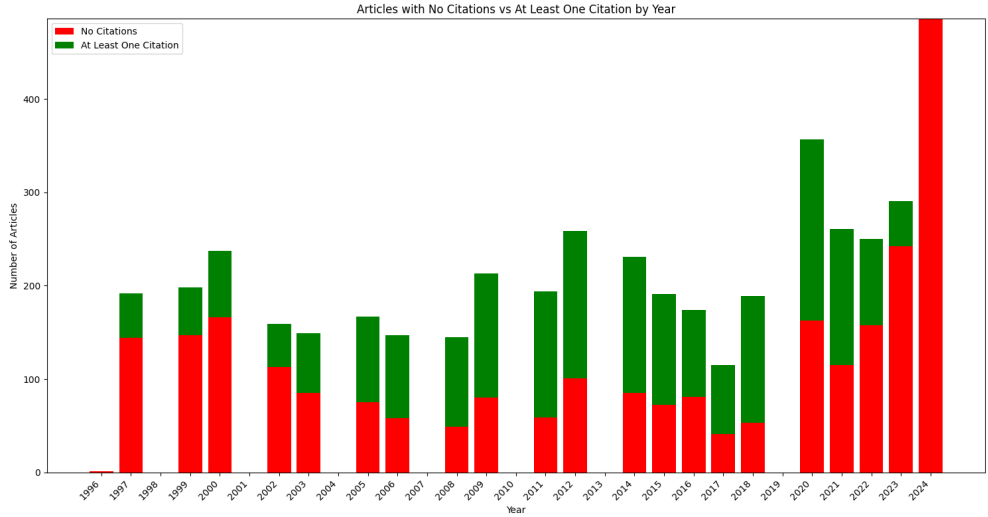}
  \label{fig:fig2}
\end{figure}

The analysis of citation data, extracted from PubMed metadata, reveals varying levels of scholarly impact across the years. Figure \ref{fig:fig2} presents a year-wise breakdown of articles that have received at least one citation versus those that remain uncited.
This visualization provides insights into the influence and reach of MIE Conference publications over time, potentially reflecting changes in research focus, quality, or dissemination strategies.
\subsection{Authorship Anomalies}
\begin{figure}[H]
  \centering
  \caption{Articles with No Authors vs At Least One Author by Year}
\includegraphics[width=0.8\textwidth]{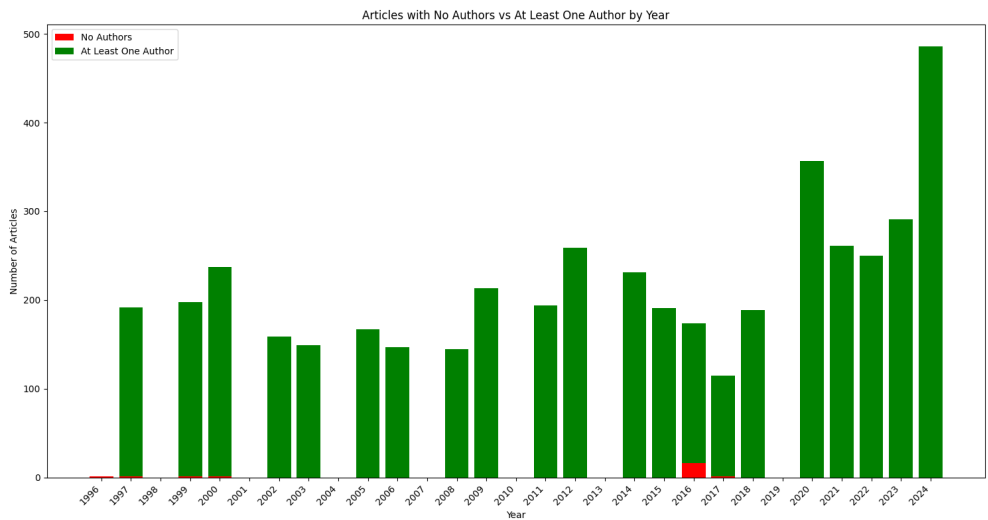}
  \label{fig:fig3}
\end{figure}

An unexpected finding in the dataset pertains to authorship information. As shown in Figure \ref{fig:fig3}, a number of articles across various years were found to have no authors attributed to them.

This anomaly appears to be related to the indexing practices of PubMed, particularly for conference proceedings. For instance, in 1996, the entire conference book was saved as a single information field in PubMed, leading to inconsistencies in individual article metadata. This observation underscores the challenges in standardizing bibliographic data for conference proceedings and highlights the need for careful data cleaning and validation in bibliometric studies.

\subsection{Affiliation Anomalities}
While parsing the affiliations, we noticed that there are some incomplete affiliations leading to inability to parse them properly. From the total number of 4606 articles, 1564 articleswere with Incomplete affiliation, which means that 33.96\% of them are incomplete.
\begin{figure}[H]
  \centering
  \caption{Articles with at least one incomplete affiliation parsing}
\includegraphics[width=0.8\textwidth]{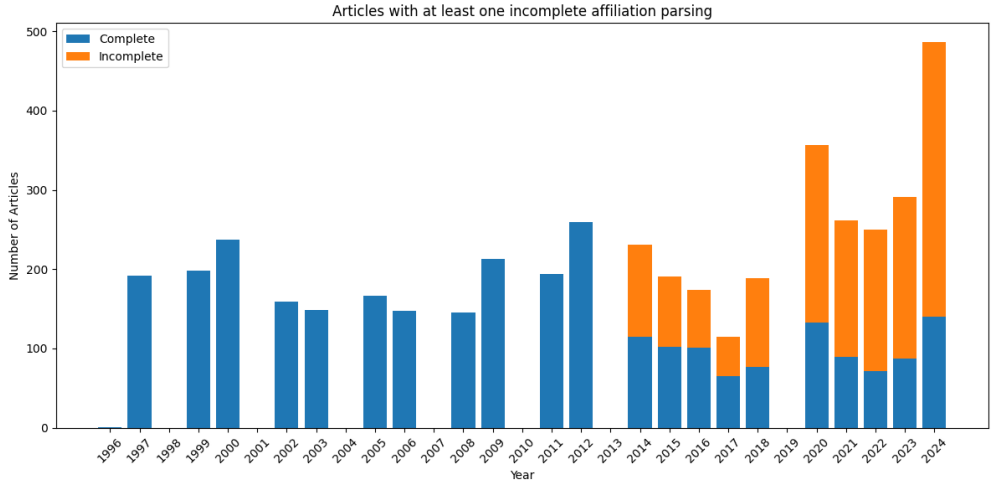}
  \label{fig:fig4}
\end{figure}

\subsection{Language of Publications}
\begin{figure}[H]
  \centering
  \caption{English vs Non-English Articles by Year}
\includegraphics[width=0.8\textwidth]{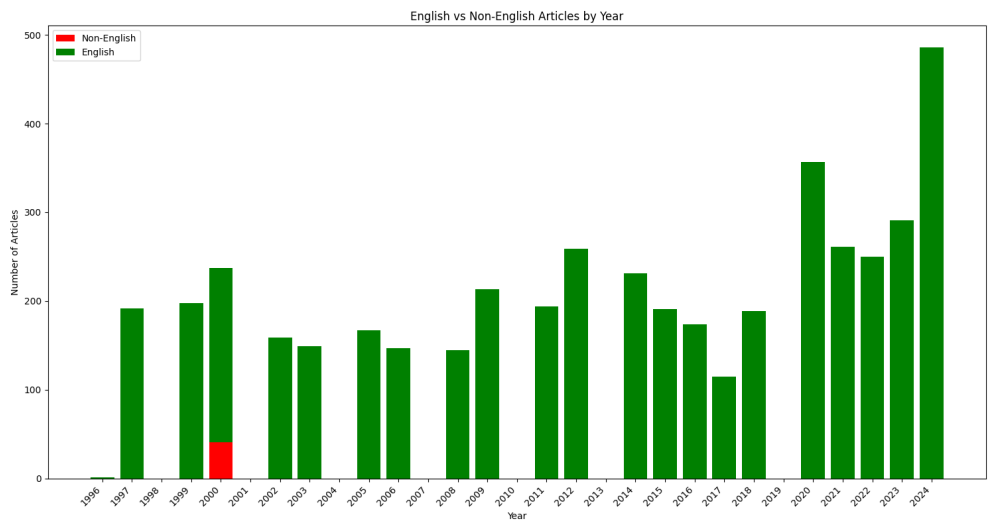}
  \label{fig:fig5}
\end{figure}

While English has been the primary language of the MIE Conference, our analysis revealed an interesting deviation in the year 2000. As depicted in Figure \ref{fig:fig5}, a small number of articles were accepted in German during this year.\\
This finding provides an intriguing glimpse into the conference's language policies and potential efforts to accommodate non-English speaking researchers. It also raises questions about the impact of language diversity on the dissemination and accessibility of research in the field of medical informatics. The detailed list of non-English publications are available as ancillary files along with this paper.

In conclusion, these results offer valuable insights into the characteristics and evolution of the MIE Conference dataset. The observed patterns in DOI assignment, citation trends, authorship anomalies, and language diversity not only inform our understanding of the dataset itself but also reflect broader trends in medical informatics research and academic publishing practices over the years.
\section{Limitations}
While the dataset presented in this article offers valuable insights into the Medical Informatics Europe (MIE) Conference proceedings, it is important to acknowledge several limitations that may impact its comprehensiveness and applicability:
\begin{enumerate}
\item \textbf{Incomplete Conference Coverage}:
\begin{itemize}
\item Not all EFMI MIE conferences are published in IOS Press or indexed in PubMed.
\item This limitation results in a partial representation of the complete MIE conference history.
\item Researchers should be aware that some conferences and their associated papers may be missing from this dataset.
\end{itemize}
\item \textbf{Temporal Constraints}:
\begin{itemize}
\item The dataset covers MIE conferences from 1996 onwards.
\item This limitation is due to the availability of data through the PubMed API and the capabilities of the TripleA program used for data collection.
\item Earlier conferences (pre-1996) are not represented in this dataset, potentially omitting historical trends and developments in the field of medical informatics.
\end{itemize}
\item \textbf{Data Source Dependence}:
\begin{itemize}
\item The dataset relies primarily on PubMed indexing and the PubMed API for data retrieval.
\item Papers or conferences not indexed in PubMed are consequently excluded from the dataset.
\item This dependence may introduce a bias towards certain types of publications or research areas that are more likely to be indexed in PubMed.
\end{itemize}
\item \textbf{Affiliation Data Limitations}:
\begin{itemize}
\item The affiliation parsing algorithm, while advanced, may not capture all nuances of institutional affiliations.
\item There might be inconsistencies or missing data in the affiliation fields, especially for older publications or those with complex multi-institutional collaborations.
\end{itemize}
\item \textbf{Topic Extraction Constraints}:
\begin{itemize}
\item The unsupervised topic extraction mechanism, while powerful, may not always accurately represent the full scope of an article's content.
\item The limitation to the top 10 topics per article might oversimplify the research focus, especially for interdisciplinary or complex studies.
\end{itemize}
\item \textbf{Citation Count Timeliness}:
\begin{itemize}
\item Citation counts are subject to the update frequency of the PubMed database and may not reflect the most current impact of the articles.
\item Newer articles may have lower citation counts due to the time lag in accumulating citations, potentially underrepresenting their importance or impact.
\end{itemize}
\item \textbf{Language Bias}:
\begin{itemize}
\item While the dataset includes a 'language' field, it predominantly contains English-language publications due to PubMed's indexing practices.
\item This may underrepresent research published in other languages, potentially missing important contributions from non-English speaking regions.
\end{itemize}
\item \textbf{Metadata Completeness}:
\begin{itemize}
\item Some fields, such as 'doi' or 'affiliation\_integration\_department', may be incomplete for certain entries, limiting the ability to perform comprehensive analyses across all dimensions of the dataset.
\end{itemize}
\end{enumerate}
These limitations should be carefully considered when using this dataset for research or analysis purposes. While the dataset provides a valuable resource for studying trends and developments in medical informatics through the lens of MIE conferences, users should be aware of its scope and potential biases. Future work could focus on addressing these limitations by incorporating additional data sources, extending the temporal coverage, and refining the data processing techniques.
\section{Conclusion}
The Medical Informatics Europe (MIE) Conference dataset presented in this article represents a significant contribution to the field of medical informatics. By compiling and enhancing data from MIE conferences since 1996, this dataset offers a unique and comprehensive resource for researchers, educators, policymakers, and industry professionals.

Key aspects of this dataset include:
\begin{enumerate}
\item \textbf{Comprehensive Coverage}: Despite some limitations, the dataset provides a broad overview of MIE conferences from 1996 onwards, offering insights into the evolution of medical informatics over nearly three decades.

\item \textbf{Advanced Data Processing}: The application of topic extraction and affiliation parsing techniques has enriched the raw bibliometric data, providing additional layers of information for analysis.

\item \textbf{Multifaceted Information}: Each entry in the dataset contains a wealth of information, including bibliometric details, extracted topics, author information, and standardized affiliation data.
\end{enumerate}
The value of this dataset lies in its potential to facilitate a wide range of analyses and applications:
\begin{itemize}
\item Longitudinal studies of trends in medical informatics
\item Bibliometric analyses to identify influential works and authors
\item Collaboration network mapping and geographical distribution of research
\item Topic modeling to track the emergence and evolution of key themes in the field
\item Curriculum development and educational planning
\item Industry and technology trend analysis
\item Identification of research gaps and future directions
\end{itemize}
While acknowledging limitations such as incomplete conference coverage and temporal constraints, this dataset nonetheless represents a valuable resource for the medical informatics community. Its comprehensive nature allows for multifaceted analyses that can provide insights into the historical development, current state, and future directions of the field.

The potential applications of this dataset extend beyond academic research. It can inform policy decisions, guide funding allocations, assist in conference planning, and help bridge the gap between academia and industry in the rapidly evolving field of medical informatics.

As the healthcare sector continues to digitize and the role of informatics in medicine grows, datasets like this one become increasingly valuable. They provide a foundation for understanding the field's trajectory and can help shape its future direction. By making this dataset available, we aim to foster collaboration, innovation, and progress in medical informatics.

Future work could focus on expanding the dataset to include more recent conferences, incorporating additional data sources to address current limitations, and developing tools to facilitate easier analysis and visualization of the data. We encourage the research community to utilize this resource and build upon it, potentially leading to new insights and advancements in the field of medical informatics.

In conclusion, this MIE conference dataset represents a significant contribution to the medical informatics community. It not only provides a window into the history and evolution of the field but also serves as a springboard for future research, collaboration, and innovation. As we continue to navigate the complex intersection of healthcare and technology, resources like this dataset will play a crucial role in shaping our understanding and guiding our progress.
\section{Ethics Statement}
The current work does not involve human subjects, animal experiments, or any data collected from social media platforms.\\
The papers accessed for this article, were published online with Open Access by IOS Press and distributed under the terms of the Creative Commons Attribution Non-Commercial License 4.0 (CC BY-NC 4.0).
\section{Data Availability}
The data is available in json format on the figshare platform\footnote[3]{ \href{https://doi.org/10.6084/m9.figshare.27174759}{https://doi.org/10.6084/m9.figshare.27174759}} with DOI \href{10.6084/m9.figshare.27174759}{10.6084/m9.figshare.27174759} licensed as CC BY 4.0. 


\end{document}